\documentclass[useAMS,usenatbib,usegraphicx,letterpaper]{mn2e}
\usepackage{times}

\newenvironment{jh}{}{}

\newcommand{\Msun}{\mbox{$M_{\odot}$}}

\newcommand{\Lsun}{\mbox{${\mathrm L}_{\odot}$}}

\defcitealias{paperI}{Paper~I}
\defcitealias{class}{Paper~II}
\defcitealias{outflows}{Paper~III}
\defcitealias{massdep}{Paper~IV}

\title[NGC1333 with SCUBA-2]{The JCMT Gould Belt Survey: SCUBA-2 observations of radiative feedback in NGC1333}
\author[Hatchell et al.]{ J.~Hatchell $^{1}$\thanks{E-mail:
hatchell@astro.ex.ac.uk},T.~Wilson $^{1}$, E.~Drabek $^{1}$, E.~Curtis$^{2}$, J.~Richer$^{2,3}$, D.~Nutter$^{4}$, \newauthor J. Di Francesco$^{5,6}$ and D. Ward-Thompson$^{7}$ \begin{jh}on behalf of the JCMT GBS consortium$^{8}$.\end{jh}\\
$^1$Physics and Astronomy, University of Exeter, Stocker Road, Exeter EX4 4QL\\
$^2$Astrophysics Group, Cavendish Laboratory, Madingley Road, Cambridge CB3 0HE\\
$^3$\begin{jh}Kavli\end{jh} Institute for Cosmology, Cambridge, Madingley Road, Cambridge, CB3 0HA\\
$^4$Department of Physics and Astronomy, Cardiff University, Queen's Buildings, The Parade, Cardiff CF24 3AA\\
$^5$National Research Council of Canada Radio Astronomy Program, 5071 West Saanich Rd., Victoria, BC, V9E 2E7, Canada \\
$^6$University of Victoria, Department of Physics and Astronomy, PO Box 3055, STN CSC, Victoria, BC, V8W 3P6, Canada\\
$^7$Jeremiah Horrocks Institute,University of Central Lancashire,Preston,Lancashire,PR1 2HE\\
$^8$\begin{jh}A complete list of survey members is given in Appendix~\protect\ref{sect:consortium} (online only).\end{jh}
}

\begin{document}

\date{Sept12}

\pagerange{\pageref{firstpage}--\pageref{lastpage}} \pubyear{2012}

\maketitle

\label{firstpage}

\begin{abstract}

We present observations of NGC1333 from SCUBA-2 on JCMT, observed as a JCMT Gould Belt Survey pilot project during the shared risk campaign when the first of four arrays was installed at each of 450 and 850 microns.  Temperature maps are derived from 450\micron\ and 850\micron\ ratios under the assumption of constant dust opacity spectral index $\beta=1.8$.  Temperatures indicate that the dust in the northern (IRAS 6/8) region of NGC1333 is hot, 20--40~K, due to heating by the B~star SVS3, other young stars in the IR/optically visible cluster, and embedded protostars.  Other luminous protostars are also identified by temperature rises at the $17''$ resolution of the ratio maps (0.02~pc assuming a distance of 250~pc for Perseus).  The extensive heating raises the possibility that the radiative feedback may lead to increased masses for the next generation of stars.
\end{abstract}


\section{Introduction}

Gas temperature plays a role in controlling star formation in dense cores through thermal support \citep{jeans02}.  The thermal Jeans length may ultimately control the sub-fragmentation of cloud cores once turbulent motions have decayed and cores become coherent \citep{goodman98_cohII, myers83_turb}.  Heating of protostellar cores suppresses sub-fragmentation \citep{bate09,offner09,urban10,chabrierhennebelle11} 
and is potentially an important factor in the formation of massive stars (along with magnetic fields, competitive accretion, and other factors; see \citealp{hennebellecommercon12} for a review). 

Observationally, several techniques exist for measuring temperatures of molecular gas, including CO excitation temperatures \citep[e.g.][]{curtis10a,ladd94} and the ammonia ladder \citep[e.g.][]{huttemeister93}.  
\begin{jh}The dust colour temperature $T_\mathrm{d}$ provides an alternative estimate of the gas temperature: at high molecular hydrogen densities $n_{H_2} > 10^{4.5} \hbox{ cm}^{-2}$, where gas-grain collisions dominate the gas heating, the dust and gas temperatures are well coupled \citep{goldsmith01}.\end{jh}  Dust colour temperatures rely on fitting dust continuum fluxes to an opacity-modified blackbody spectrum, 
a method which is being applied very successfully to make temperature and column density maps of nearby star-forming regions from Herschel data based on fits from 160~\micron\ to 500~\micron\ \citep{andre10}. 

In this paper we apply dust temperature methods to 450\micron\ and 850~\micron\ submillimetre (submm) data from SCUBA-2 on the James Clerk Maxwell Telescope \citep[JCMT; ][]{holland06}, using a similar technique to that developed for SCUBA \citep{kramer03}.  There are benefits in using long-wavelength mm/submm data for dust temperatures (as long as the data points are not entirely on the Rayleigh-Jeans tail).   When temperature varies along the line-of-sight, dust continuum fluxes are emission-averaged and hence weighted towards the hottest dust.  Small fractions of hot dust can dominate the mid-infrared (mid-IR) even though they contain only a small fraction of the total dust mass, as is evident from protostellar spectral energy distributions \citep[e.g.][]{enoch09}.  For sources with a simple geometry, this temperature variation issue can be circumvented by radiative transfer modelling, but this is not a realistic option for complex star-forming regions.   Also, in high column density regions, the dust emission at mid-IR wavelengths can become optically thick, weighting the results towards the front side of the cloud and low-opacity photon escape routes.  The $14''$ beam full width half maximum (FWHM) of JCMT at 850~\micron\ provides an improvement in angular resolution over Herschel at 500~\micron\ ($37''$~FWHM), which is advantageous in dense clusters and when determining the relative location of heating sources and cores.

Here we apply the dust colour temperature method to the young cluster NGC1333 in Perseus.  NGC1333 is a young cluster well studied in the mm/submm \citep{sandellknee01,hatchell07a,kirk07,enoch06,curtis10a,curtis10b}; for general background see \citet{walawender08}.  

\section{Method}


\subsection{Observations and data reduction}
\label{sect:observations}

NGC1333 was observed on 
February~21--22~2010 and March~4th and 7th~2010 as a JCMT Gould Belt Survey \citep{JCMTGB} pilot project in the SCUBA-2 shared risk (S2SRO) campaign, during which only one array out of the final four was available at each of 850\micron\ and 450~\micron\ \citep{holland06}. 
 These observations were among the first made with SCUBA-2.
The weather conditions were good, JCMT Band~2 with 225~GHz opacity $\tau_{225} = 0.05\hbox{--}0.07$.  Two overlapping $15'$ diameter circular regions \citep[`PONG900' mapping mode;][]{jenness11} were targetted, with map centres offset $10'$ to the north and south from $\rmn{RA}(2000)=03^{\rmn{h}}~29^{\rmn{m}}~00\fs0$,
$\rmn{Dec.}~(2000)=31\degr~18\arcmin~00''$ .
These fully-sampled $15'$-diameter regions were surrounded by $4'$ borders that were scanned at lower sensitivity by decreasing numbers of the array pixels.  

Molecular lines contribute to the signal in the SCUBA-2 850~\micron\ continuum bandwith, with the most significant contamination at 345~GHz due to the CO~$J=3\hbox{--}2$ line \citep{drabek12}.  The CO contamination was removed during the 850~\micron\ reduction using the HARP~$^{12}$CO~$3\hbox{--}2$ integrated intensity map \citep{curtis10a} converted to a CO contamination map using a contamination factor of 0.66~mJy/beam/(K~km~s$^{-1}$) appropriate to Band~2 weather with an average $\tau_\mathrm{225} = 0.06$ \citep{drabek12}.  The 850~\micron\ map was re-reduced in June~2012 using the development version of the data reduction package SMURF \citep[][version of  2nd~May~2012]{jenness11} with astronomical signal (AST) masking based on earlier versions of the reductions and gridding to $6''$ pixels.   The CO contamination map was injected during the 850~\micron\ reduction process as a fake source with a negative multiplying factor of $-1/\mathrm{FCF}$ where FCF is the flux conversion factor (see below).  In this way, the CO emission was subtracted from the timeseries fluxes during the reduction, matching the spatial filtering inherent in SCUBA-2 map reconstruction and resulting in an 850~\micron\ map which is corrected for CO line contamination.  The 450~\micron\ map was re-reduced in May~2012 using the same SMURF version, 
with AST masking, without CO contamination removal, 
and gridded to $4''$ pixels.  Both the 850~\micron\ and 450~\micron\ maps are insensitive to emission on scales greater than the on-sky footprint of the array ($4'$ for the single-array S2SRO)  due to common-mode subtraction of atmospheric emission across the array.

Maps were calibrated by multiplying by flux conversion factors (FCFs) of $653 \pm  49$ Jy/beam/picoWatt and $511 \pm 31$ Jy/beam/pW at 850~\micron\ and 450~\micron\ respectively, derived from the six observations of the submm calibrator CRL618 \citep{dempsey12}. 
The average beam FWHM were measured to be $14.4'' \pm 0.15''$ (850~\micron) and $9.8'' \pm 0.23''$ (450~\micron) consistent with the observatory standards of $14.2'' / 9.4''$.

Of the 18 maps observed, 12 were selected for mosaicking, six each for the north and south centres, with six rejected due to telescope tracking errors.  The resulting mosaics have pixel-to-pixel $1\sigma$ noise levels of 7.8 and 130 mJy per $14.4''/9.8''$ beam on $6''/4''$ pixels at 850~\micron/450~\micron\ respectively (an average over the north and south regions, which have noise levels +4\% higher and lower respectively), roughly a factor two deeper than SCUBA \citep{sandellknee01,paperI}.  Calculating the equivalent point-source sensitivity at a resolution equal to the beam area, these observations meet the JCMT Gould Belt Survey 3~mJy target  at 850~\micron\ while achieving 50~mJy point source sensitivity at 450~\micron.  
 The central, overlap region of the map has a still lower noise level by roughly $\sqrt{2}$.  The pixel-to-pixel noise estimates do not completely represent the noise levels in larger-scale fluctuations which were measured from histograms of the pixel values to be 14~mJy/beam and 150~mJy/beam respectively.



\subsection{Ratio and temperature maps}

For optically thin emission, the flux ratio between the two wavelengths 450~\micron\ and 850~\micron\ depends solely on the dust opacity and temperature.  The dust opacity is standardly parameterised at long wavelengths as a power law $\kappa  \propto 
\nu^\beta$ \citep{hildebrand83}.  The 450/850 flux ratio is then
\begin{equation}
{S_{450} \over S_{850}} = \biggl({850\over
  450}\biggr)^{3+\beta}\biggl({\exp( hc/\lambda_{850} kT_\mathrm{d}) - 1 \over \exp(hc/\lambda_{450}kT_\mathrm{d})
  - 1}\biggr) \label{eqn:beta}
\end{equation}
where $T_\mathrm{d}$ is the dust temperature and $\lambda_{850}$ and $\lambda_{450}$ are 850~\micron\ and 450~\micron\ wavelengths respectively.

Higher 450/850 flux ratios can be produced either by higher dust temperatures, a higher opacity index $\beta$, or both.  Separating the two requires data at additional wavelengths, e.g. a joint analysis with the Herschel 250~\micron\ and 500~\micron\ datasets, which is planned for future work beyond the scope of this Letter.  Higher dust temperatures are caused by heating, whereas grain growth can change $\beta$ by raising the typical sizes and absorption wavelengths of grains. $\beta$
typically lies in the range 1 to 2, depending on the type and
clumpiness of grains.  Increases in $\beta$ have been observed in protoplanetary disks \citep[e.g.][]{ubach12} and low-density clouds \citep{martin12} but to date there is little knowledge of changes in pre/protostellar cores, a situation that should improve rapidly with availability of SCUBA-2 and Herschel data.  There is a suggestion that icy mantles increase towards the inner parts of the Class~0 source B335 though not enough to significantly affect $\beta$ \citep{doty10}.


The flux ratio map was created from the 450~\micron\ and CO-subtracted 850~\micron\ maps, which were scaled to Jy/pixel with any negative fluxes removed. The 450~\micron\ map was convolved with a 2-dimensional Gaussian representing the 850~\micron\ $14.4''$ primary beam, and the 850~\micron\ map similarly convolved with a 450~\micron\ $9.8''$ primary beam to match angular resolution.  The cross-convolution procedure produced more consistent matched beams than simply convolving the 450~\micron\ map to the 850~\micron\ resolution and was needed for the S2SRO data to avoid artefacts in the ratio map.  
The flux ratio map was created by dividing the resolution-matched, regridded 450~\micron\ and 850~\micron\ maps with values in each map below $5\sigma$ blanked.  For a fixed value of the dust opacity $\beta$, Equation \ref{eqn:beta} was used to convert the ratio map into a temperature map, with an additional cut of temperatures with variance $>30$~K applied.  The effective angular resolution of the ratio and temperature maps is $17.4''$.  The 
systematic uncertainty in the ratio maps due to calibration at each wavelength is 15\%.  The large-scale filtering inherent in S2SRO maps adds edge effects and systematic offsets to the ratios which will be quantified by future modelling.  These effects are smallest where bright compact structures dominate the fluxes, as is the case for most of NGC1333. 

\begin{figure*}
\includegraphics[width=18cm]{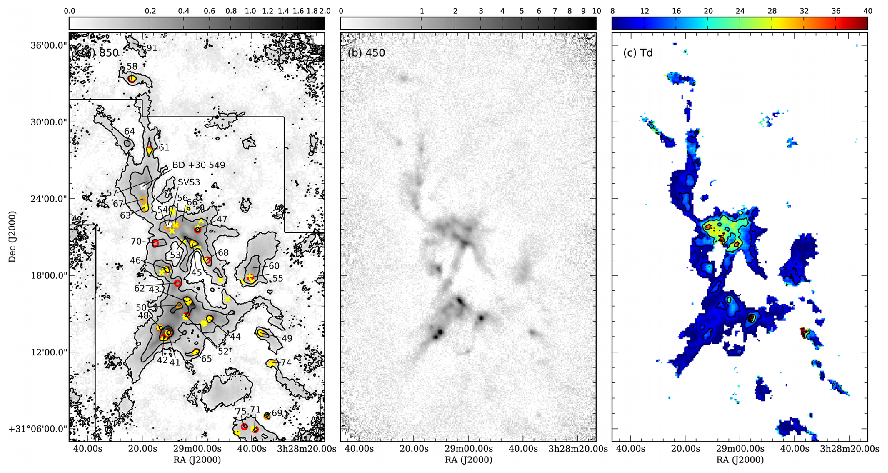}
\caption{Left to right: (a) CO-subtracted S2SRO 850~\micron\ map of NGC1333.  Contours are at 0.02,0.1,0.5, 2.5 Jy/beam.  The black line marks the CO subtraction region.  Labels identify SCUBA sources \protect{\citep{paperI}} and the two B~stars.  Markers identify embedded protostars from \protect{\citet{hatchell07a}} (red circles for Class~0,  orange circles for Class~I), \protect{\citet{jorgensen06a}} (yellow $+$) and \protect{\citet{gutermuth08}} (yellow $\times$). (b) S2SRO 450~\micron\ map in Jy/beam.  (c) Dust temperature $T_\mathrm{d}$ map at $\beta=1.8$ in units of Kelvin.  Contours are at $T_\mathrm{d} = 12$, 20, and 30~K.}
 \label{fig:ngc1333}
\end{figure*}

\section{Results}
\label{sect:results}

The S2SRO 850~\micron\ map of NGC1333 is presented in Fig.~\ref{fig:ngc1333}(a) as a finding chart.  SCUBA cores are labelled with their HRF2005 numbering  \citep[hereafter HRF$nn$]{paperI,hatchell07a} and protostars identified from {\em Spitzer} mid-IR  detections are marked \citep{gutermuth08,jorgensen06a}.  Further identifications of sources in NGC1333 can be found in \citet{sandellknee01}. Figure~\ref{fig:ngc1333}(b) shows the 450~\micron\ map of this region.  The 450~\micron\ dataset from S2SRO is sensitive and stable enough for the filamentary structure to be seen clearly, as it is not dominated by the large-scale noise artefacts which limited analysis of the SCUBA 450~\micron\ data \citep{sandellknee01,paperI}.

%
%



In Fig.~\ref{fig:ngc1333}(c) we plot the dust temperature in NGC1333 assuming a constant dust opacity spectral index of $\beta=1.8$.  This choice of $\beta$ is consistent with the popular OH5 dust model \citep[$\beta=1.85$; ][]{oh94}, with recent results for dense cores from Planck and Herschel \citep{juvela11,stutz10}, and with comparisons of SCUBA and near-infrared opacities \citep{shirley11}.   Although variation in the 450/850 ratio may also be due to variation in $\beta$ (to be investigated in future work), as NGC1333 contains several luminous protostars and a reflection nebula, it is reasonable as a first approximation to assume that the variation is dominated by temperature.  \begin{jh}A higher value for $\beta$ would result in lower temperatures for the same ratios: temperatures of 10~K, 20~K and 40~K at $\beta=1.8$ fall to 8.5~K, 14~K and 20~K at $\beta=2.2$. \end{jh}

Filaments and ambient cloud material generally show dust temperatures of 10~K (corresponding to 450/850 ratios of around 4; ).  These temperatures are typical of the cold dense interstellar medium \citep{evans01,shirley02}.  Towards the protostars, 
beam-averaged temperatures rise to 12-15~K, consistent with SCUBA-based models of low-mass Class~0 and 1 protostars \citep{shirley02,young03} and with the Herschel analysis of CB244 \citep{stutz10}.  Some of the more luminous protostars show notably higher dust temperatures as discussed below.   The temperature does not, however, rise towards all protostars identified by Spitzer mid-IR detections or outflows \citep{jorgensen06a,hatchell07a,gutermuth08,hatchell07b,hatchell09}.   
Low-luminosity protostars have similar 
temperatures of 10--15~K to starless cores. The falling 450/850 ratios towards the map edge are almost certainly an artefact of the filtering of the large-scale background in the S2SRO map reconstruction, and not a physical effect due to falling $\beta$ or temperature.  Edge reconstruction should improve with data from the full SCUBA-2 array, which is sensitive to spatial scales a factor of two larger.

The most obvious feature in the temperature map is the warm northern region containing cores HRF45, 
47, 54, 56, and 66 
that shows widespread dust temperatures above 20~K.  This region lies to the south of the optical cluster, and its dust is presumably heated by the stars in the optical cluster as well as the embedded protostellar population.   The optically revealed cluster stars are dominated by the B5~star SVS3 \citep{strom76,straizys02} which lies on the edge of the dust to the northeast of HRF54, as marked in Fig.~\ref{fig:ngc1333}(a).   SVS3 is a 138~\Lsun\ binary with an F2 companion \citep{connelley08}; the optically-visible NGC1333 reflection nebula and associated infrared source \citep[IRAS 8;][]{jennings87} is evidence its proximity to the dust cloud.  The other main heating sources are embedded protostars.   Core HRF45 is heated by SSTc2d~J032901.6+312021 \citep{evans09} as supported by its high temperature and identification as IRAS~6 \citep{jennings87}.  Protostar SSTc2d~J032907.8+312157 in HRF56 is also classified as a 27\Lsun\ luminosity source \citep{gutermuth08,evans09}.  This luminosity is likely to be partially due to accretion and partially reprocessed flux from SVS3.  The dust at this position is warm ($\sim 20$~K), though not as warm as neighbouring HRF54 heated by SVS3 to form IRAS~8.   The ridge containing HRF66 has elevated temperatures; a protostar embedded in HRF66 could explain some of this heating but a second source would also be necessary to heat the ridge to the southwest.  Sources HRF47 and 53 are also warm ($\sim20$~K) but do not show temperature peaks. Core HRF47 contains a Spitzer-identified protostar, J032859.56+312146.7 \citep{jorgensen06a,evans09}, but the closest Spitzer detection to HRF53 is SSTc2d~J032905.18+312036.9 to the southeast where the dust temperature rises \citep{gutermuth08,evans09}.  

An insight into the location of the heating sources can be made by comparison with $^{12}$CO excitation temperatures $T_\mathrm{ex}(\mathrm{CO})$ \citep[][~Fig. 13]{curtis10a}.  The $^{12}$CO traces warm gas on the front side of the cloud (as the $^{12}$CO is generally optically thick) but the submm dust emission is optically thin and traces heating through the cloud.  
Towards SVS3, the $^{12}$CO temperature peaks sharply at over $40$~K indicating heating of the front of the cloud.  The dust temperature rise is more extensive, suggesting additional heating from embedded sources or sources behind the cloud, including SSTc2d~J032905.18+312036.9. 
Towards the HRF45 and 47 cores, the dust and $^{12}$CO excitation temperatures are similar ($\sim40\hbox{--}45$~K), indicating that the heating sources lie near the front side of the cloud.

Further to the north, another B star, BD~$+30$~549 or NGC1333~IRAS~9 classified as B8 \citep{jennings87}, lies close to HRF57 and 67 in projection which may explain raised dust temperatures of 15-17~K in these northern filaments.  A significantly higher CO temperature, $T_\mathrm{ex}(\hbox{CO})$ over 40~K,  suggests that BD $+$30~549 lies in front of the cloud.  Several protostars in other parts of the cloud are identifiable by their raised 450/850 ratios and temperatures.  NGC~1333 IRAS~7 (HRF46) shows raised  $T_\mathrm{ex}(\hbox{CO}) \sim 35$~K and raised $T_\mathrm{dust} \sim 20$~K to the northeast of the dense core.  This could be due to the embedded source lying towards the front side of the cloud, or more likely in this case, hot CO lining a protostellar outflow cavity opening to the northeast \citep{curtis10b}.

Deeply embedded sources or sources behind the cloud show the opposite effect with $T_\mathrm{d} > T_\mathrm{ex}(\hbox{CO})$.  These include the SVS13 ridge (HRF43) and NGC~1333 IRAS~2 (HRF44).  The HRF49 condensation shows a particularly strong dust temperature gradient from 10K in the southwest to over 40K at the northeast head of the condensation, heated by infrared source SSTc2d~J032837.1$+$311331 \citep{gutermuth08,enoch09}.  $^{12}$CO excitation temperature maps \citep[][Fig.~13]{curtis10a} show no evidence for raised gas temperatures in this region indicating that the front of the core from which the optically-thick $^{12}$CO escapes is cold and the star and the hot dust lie behind the dense core.  
%


\section{Discussion}
\label{sect:discussion}


\begin{jh}Our observations show that star formation has raised 450/850 ratios in NGC1333.  The absolute values of temperature in Fig.~\ref{fig:ngc1333} depend on our assumption of $\beta=1.8$, and increases in $\beta$ may account for some of the 450/850 ratio rises towards dense cores.  However, an explanation based on opacity alone would require extreme values of $\beta\sim 3$ towards the reflection nebula and luminous SVS13, whereas dust heating provides a simple explanation for ratio rises in these regions.\end{jh}

According to theory, heating increases thermal pressure which acts to stabilise cloud material against further fragmentation \citep{jeans02}.  The thermal Jeans length and Jeans mass scale as $\sqrt{T}$ and $T^{3/2}$ respectively.  
Cores may be heated internally, potentially suppressing fragmentation resulting in more massive stars, or externally by earlier generations of stars, as recently investigated in G8.68$-$0.37 \citep{longmore11}.  \begin{jh}In NGC1333, \end{jh}there is no evidence that the northern region has fragmented with larger core separations than the rest of \begin{jh}the region\end{jh} to date, presumably because the $1'\hbox{--} 1.5'$ core separation scales have been set by turbulent, rather than thermal, support \citep[e.g.][]{chabrierhennebelle11}. 
The local temperature rise in NGC1333~N due to the existing optical cluster and embedded protostars may, however, now inhibit the formation through fragmentation of further gravitational cores within the northern region. At the average molecular hydrogen density of NGC1333~N, $n_{\mathrm{H}_2} \simeq 4 \times 10^4\hbox{ cm}^{-3}$, the Jeans mass increases from 0.5 to 4~\Msun\ as temperature rises from 10 to 40~K.   If no further cores form, the existing protostars will accrete the remaining gravitationally bound gas (minus a fraction removed by outflows).   The existing cores will produce more massive stars than they would have done in the absence of external heating.  Hence the radiative feedback from the existing cluster may influence the development of the next generation of stars.

There is possible evidence in the data here of the effect of external heating in promoting massive star formation in core HRF56.  The embedded protostar SSTc2d~J032907.8+312157 lies only 0.05~pc in projection from the B5 star SVS3 and on the edge of the optical nebula.  If its current luminosity of 27\Lsun\ \citep{evans09} is internally generated (rather than reprocessed emission from SVS3), and with a few~\Msun\ envelope still to accrete, HRF56 could be forming a second B star.  Its lack of fragmentation could be influenced by heating from nearby SVS3.  The nearby HRF54 core may be affected similarly in the future.  We agree with \citet{longmore11} that this is a somewhat unsatisfactory circular argument as it requires a pre-existing massive star (SVS3) to provide the feedback, rather than a cluster of low-mass stars.  
Further to the south, HRF45/IRAS~6 has a current luminosity of 15\Lsun\ and a few~\Msun\ envelope so could also reach B-star status.  This source, however, does not lie close to any such luminous star and a double radio detection indicates that it has already fragmented \citep{rac99}.  Any further stars formed are likely to be typical AFGK-type cluster members since the remaining cores are low mass, at most a few~$\Msun$ each.   (Note that the total mass in NGC1333~N is only $\sim 20\Msun$, assuming an average dust temperature of 20~K. Previous core mass calculations assumed fixed temperatures \citep{hatchell07a,kirk07,sandellknee01} or temperatures from IRAS \citep{sandellknee01}.  A future step will be to recalculate the masses of the NGC1333~N cores based on the dust temperatures and fluxes from the full SCUBA-2 Gould Belt survey data, which will have better spatial response and stabler calibration than the S2SRO data presented here).   


Finally, in HRF54/IRAS8, radiative feedback by SVS3 may be promoting formation of an early B-type star with potentially devastating consequences for NGC1333 through widespread dissociation of the molecular gas and a halt to further star formation activity in the region.  Possibly, the high-mass stars currently missing from the IMF of Perseus and other young low-mass star-forming regions \citep{evans09} only form once radiative feedback from the existing clusters reaches a sufficiently high level.  A further consequence is that massive stars are likely to form relatively late in molecular clouds, in the centre of clusters where the gas heating is greatest.

\section{Conclusions}
\label{sect:conclusions}

SCUBA-2 S2SRO observations show significantly raised dust temperatures around several luminous protostars in NGC1333, including SVS13 and IRAS~2.   \begin{jh}Assuming a dust opacity spectral index $\beta=1.8$,\end{jh} the region on the southern edge of the optical reflection nebula is extensively heated to temperatures $>20$~K, reaching $40$~K close to the B~star SVS3 and the embedded protostar IRS~6.  A significant fraction of the energy input is from the optical cluster, and specifically the B~star SVS3 illuminating the reflection nebula.  Heating by the existing cluster is potentially biassing the star formation in the north of NGC1333 in favour of more massive stars. 

NGC1333 will be observed again with the fully-commissioned SCUBA-2 as part of the JCMT Gould Belt Survey.

\section*{Acknowledgements}

JH would like to thank all the staff at JAC, UKATC, and elsewhere who have worked so hard to make SCUBA-2 a reality, the JCMT Gould Belt Survey members (see Appendix~\ref{sect:consortium})  for planning the SCUBA-2 survey, and Samantha Walker-Smith for help with protostellar core identifications.

\bibliographystyle{mn2e}
\bibliography{perseus}

\appendix
\section{The JCMT Gould Belt Survey consortium}
\label{sect:consortium}


The full members of the JCMT Gould Belt Survey consortium are 
P. Bastien, D. Berry, C. Brunt, J. Buckle, H. Butner, A. Cabral, P. Caselli, S. Chitsazzadeh, H. Christie, A. Chrysostomou, \begin{jh}S. Coud{\'e},\end{jh}, R. Curran, M. Currie, E. Curtis, C.J.~Davis, W. Dent, E. Drabek, J. Di Francesco, M. Fich, J. Fiege, \begin{jh}L. Fissel, \end{jh} P. Friberg, R. Friesen, G. Fuller, S. Graves, J. Greaves, J. Hatchell, M. Hogerheijde, W. Holland, T. Jenness, D. Johnstone,  G. Joncas, J.M.~Kirk, H. Kirk, L.B.G.~Knee, B.C.~Matthews, H. Matthews,\begin{jh} J.C.~Mottram,\end{jh} D. Nutter, J. Patience, J.E.~Pineda, C. Quinn, J. Rawlings, R. Redman, M. Reid, J. Richer, E. Rosolowsky, S. Sadavoy, C. Salji, G. Schieven, N. Tothill, H. Thomas, S. Viti, S. Walker-Smith, D. Ward-Thompson, G.J.~White, J. Wouterloot, J. Yates, \begin{jh}and \end{jh} M. Zhu.
\end{document}